# Power, Depletion and Energy Quality Model of Thermo-industrial Civilization

J.L.V. Lewandowski[1]

Institute For Collapse Studies

Rue Jean-Jacques Rousseau

Paris, Cedex 75001, FRANCE

## Abstract

The current thermo-industrial civilization is critically dependent on fossil fuel energy sources. An intuitive model capturing the interplay between economic activity, physical power consumption, depletion and energy quality is presented.

## Introduction

The strong correlation between economic activity and power (energy per unit of time) is well-established. The sources of power (electricity, natural gas, oil, coal, and renewable sources like biomass, solar, and wind) enable, for example, heating, cooling, transportation, lighting, chemical manufacturing, steel production, aluminum refining and cement manufacturing.

Although there many definitions of energy ('capacity for doing work' etc.), we prefer to consider what energy "does": *any transformation from a given state to a different state requires energy* (potential, kinetic, thermal, electrical, chemical, nuclear, etc.). Economic activity can then be measured by the energy involve in all (economic) transformations per unit of time; therefore, the overall power consumed by the economic superorganism is a useful metric to quantify its overall activity. This metric is not expressed in currencies (euro, yuan, …) by in Watts (the unit of power) or Terawatts ($10^{12}$ Watts or 1000 billions Watts). In 2024 the world power consumption was 17.7 Terawatts [1].

## Derivation of the power cliff model

Figure 1 depicts the key component of the workflow of the power cliff model. The first step is to determine the dynamical evolution of the cumulative extracted *raw* energy from which we can compute both the raw power and the energy quality at any given time. Since energy is required to extract energy, the net power, which is the useful power feeding the economic superorganism, is smaller than the raw power.

---

[1] <u>InstituteCollapseStudies@gmail.com</u>

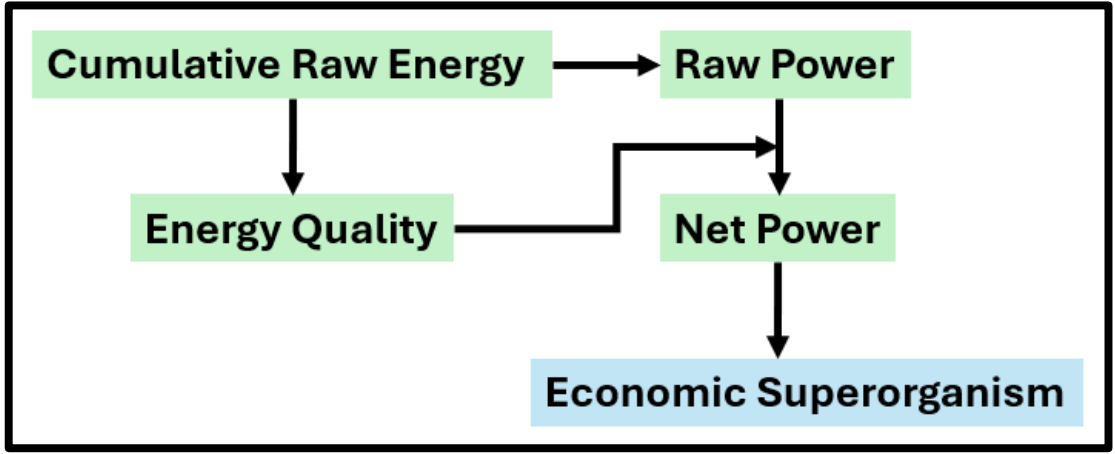

Figure 1: The power cliff workflow.

The world economic activity is denoted $S(t)$; as noted above, its units are those of power. Its dynamic evolution is

$$\frac{dS}{dt} = \gamma(x)S \tag{1}$$

where the growth rate depends on a variable $x$ which we introduce shortly. The choice of an exponential, as in Equation (1), is justified by the fact that the human population has grown at an exponential rate and that power capita has steadily increased over time. The power dissipated through economic activity must originate from an input source which we denote as $P_{eff}$ (for 'effective power'—see discussion below); it is reasonable to assume that the overall growth rate must depend on the ratio $P_{eff}/S$. The variable $x$ in Equation (1) is defined as

$$x = \eta \frac{P_{eff}}{S} \tag{2}$$

where $\eta$ is a unitless constant. Indeed, the greater the effective power per unit of economic activity the faster the growth rate. However, even if the available effective power is extremely large, the growth rate cannot exceed an upper bound, which we denote $\gamma_M$; after all, the active 'agents' in the economic superorganism are humans and their socio-politico-economic structures cannot evolve arbitrarily fast. At the same time, a minimum of (effective) power is necessary to simply *maintain* the economic superorganism: for example, even if the economy does not grow, transportation, heating, manufacturing production, lighting, chemical manufacturing, steel production, aluminum refining and cement manufacturing must continue. Therefore, there is a critical value, denoted $x_c$, for which the economic growth rate vanishes (and below which, recession or depression ensues)

$$\gamma(x_c) = 0 \tag{3}$$

The exact form of the growth rate is not known *a priori*. Here we tentatively propose that, given an increment $\Delta x$, the corresponding increment for the growth rate is

$$\Delta \gamma = -\mu(\gamma - \gamma_M)\Delta x \tag{4}$$

with positive constant $\mu$. Equation (4) suggests that, as the effective power becomes larger, the increment in growth rate declines and then becomes zero; this is consistent with the above remark that the economic superorganism cannot grow at an arbitrarily fast pace. Integrating Equation (4), and using the boundary condition Equation (3), we find

$$\gamma(x) = \gamma_M[1 - \exp(-\mu(x - x_c))] \qquad (5)$$

Figure 2 depicts the model economic growth rate as a function of the ratio of effective power to economic activity.

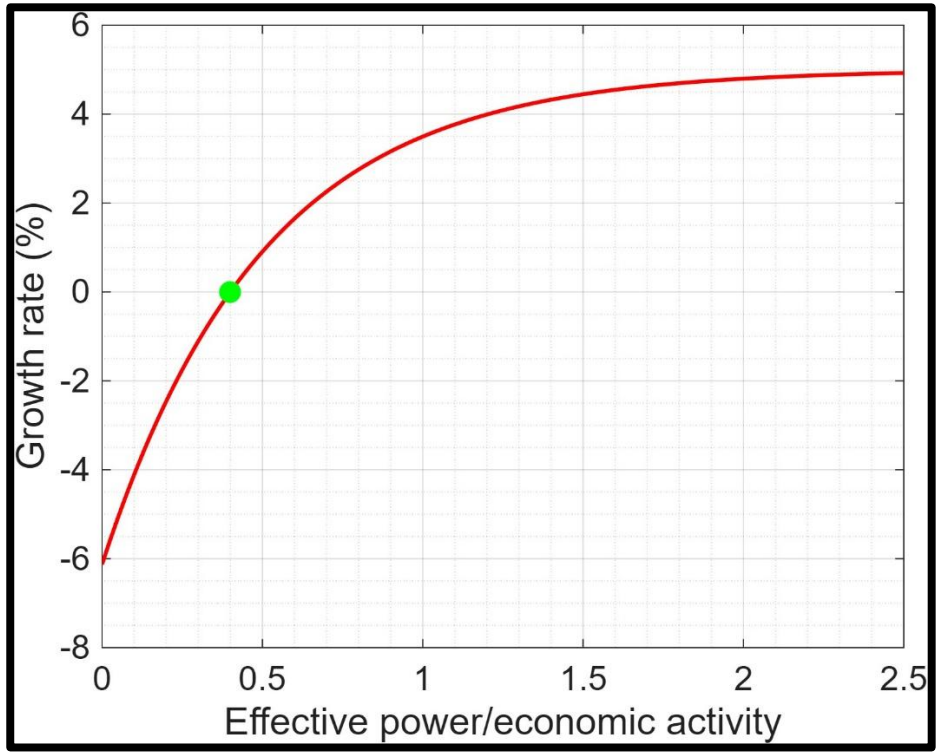

Figure 2: Economic growth rate as a function of the ratio of effective power to economic activity. The fastest growth rate is chosen to be 5%; other parameters are $\mu = 2$ and $x_c = 0.4$. The green dot shows the point at which all effective power is used to maintain the economy (that is, no growth nor contraction).

We now turn our attention to the cumulative raw energy available for global economic activity. As shown in Figure 3, the main energy sources are fossil fuels (oil, natural gas, coal). Although the wind and solar energy sources are being added at a rapid pace over the past decade, they represent a small fraction of the total global primary energy consumption. Further, the manufacturing of wind turbines and solar panels require significant amounts of fossil fuels. As a first approximation, we restrict our attention to fossil fuels.

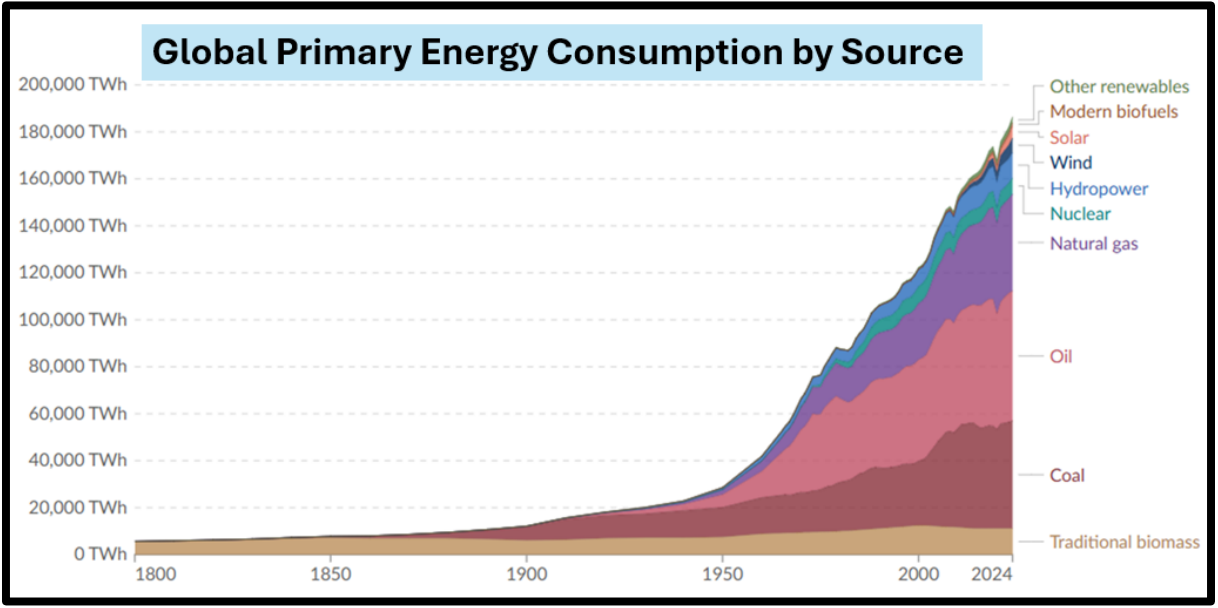

Figure 3. Global primary energy consumption by source (adapted from Reference [2]).

Since fossil fuels are non-renewable, finite resources, depletion and scarcity will ultimately occur. Peak extraction of oil, natural gas and coal are, of course, of critical importance. Maggio and Cacciola in a 2012 paper [3] proposed a variant of multi-cyclic Hubbert approach to determine the peak extraction years for oil, natural gas and coal. Their estimated peak years were 2009–2021 for oil, 2024–2046 for gas and 2042–2062 for coal. This is approximately in line with what the International Energy Agency (IEA) reported in 2025 [4].

Instead of considering each fossil fuel energy source separately, we aggregate them into a single energy source. The cumulative (raw) energy is then

$$E(t) = \alpha_o\, E_o(t) + \alpha_{ng} E_{ng}(t) + \alpha_c E_c(t)$$

where $E_o, E_{ng}, E_c$ are the cumulative energy contribution from oil, natural gas and coal, respectively. The factors $\alpha_o, \alpha_{ng}$ and $\alpha_c$ are relative fractions that, in general, depend on time. We denote $E_T$ has the total raw energy that is commercially viable [the meaning of 'commercially viable' is defined below using the concept of Energy Return on Investment (EROI)]. The evolution of cumulative raw energy extracted is governed by

$$\frac{dE}{dt} = \omega E \left[ 1 - \left(\frac{E}{E_T}\right)^\beta \right] \qquad (6)$$

where $\beta$ is a parameter of the order of unity and $\omega$ is a nominal extraction rate. A similar model (with $\beta = 1$) was suggested by M.K. Hubbert in 1956 for oil extraction with some notable success [5]. In fact, this exponent is a proxy for the technological level: the larger $\beta$, the more advanced the technology. Note, however, that technology *does not create* resources: the total raw energy $E_T$ that is commercially viable is fixed [a condition imposed by geo-physical constraints (in particular, the second Law of Thermodynamics) *not* human ingenuity]. Figure 4 compared the time evolution of raw power for 3 different values of the technological proxy, $\beta$.

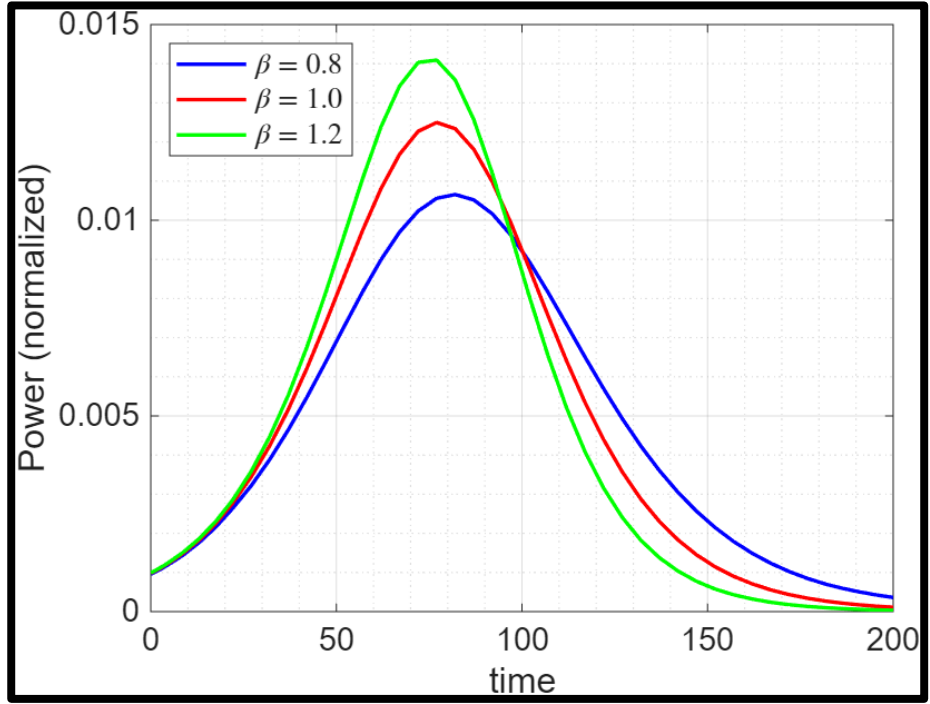

Figure 4. Raw power (normalized to $\gamma_M E_T$) as a function of time for 3 different values of the exponent $\beta$ [see Equation (6) in the main text]. The areas under each curve are the same, reflecting the fact that the stock of economically viable energy resources is fixed. The exponent is a proxy for the level of technological prowess; higher values bring an earlier, higher peak power followed by a more abrupt fall thereafter. Therefore to extract non-renewables resources (such as fossil fuels) at the fastest possible rate (as so many countries have done) is likely to bring significant economic hardships once the peak of raw power is passed.

Upon solving Equation (6) one can compute the raw power as

$$P(t) = \frac{dE}{dt} \qquad (7)$$

Next we investigate the issue of energy quality (Figure 1). The importance of energy consumption allowing societies to thrive is well established and prospects of energy needs are well known throughout the scientific literature. However, lesser discussions persist on the future availability of energy for current industrial economies, a crucial indicator for development. One common indicator is the Energy Return on Investment (EROI) which was popularized in the mid-1980s by Hall, Cleveland and Kaufmann [5]. The EROI is the ratio of the amount of usable energy (also known as exergy) delivered from a particular energy source to the amount of exergy used to obtain that energy resource (this applies not only to fossil fuels energy sources by also to renewable energy sources); we use the symbol $\theta$ to denote EROI

$$\theta = \frac{E_{out}}{E_{in}} \qquad (8)$$

As it is evident from the above equation, an energy source with a large EROI provides a large benefit to the thermo-industrial civilization (Figure 5). If the EROI drops below unity, the energy source is not economically viable as the energy cost becomes larger than the energy benefit. From this discussion it follows that the relevant quantity is the *net* energy which is given by

$$E_{net} = E_{out} - E_{in} = E_{out}\mathbb{Q}(\theta) \tag{9}$$

where, for lack of a better term, $\mathbb{Q}(\theta) \equiv 1 - 1/\theta$ is called the quality factor.

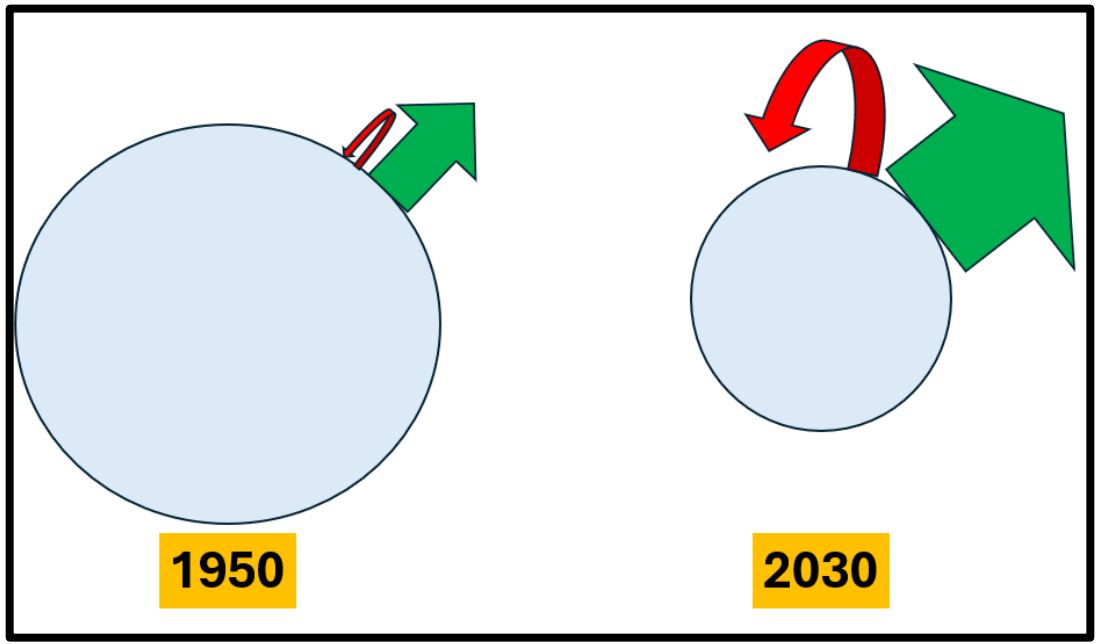

Figure 5: Illustration of the Energy Return on Investment (EROI) for fossil fuels. The area of the blue disk is proportional to the amount of fossil fuels economically extractable (economically viable). The green arrows represent the net energy whereas the red arrows represent the energy expended. The double bind of depletion becomes immediately apparent: since the beginning of the Great Acceleration (circa ~ 1950) the pool of fossil fuels economically extractable has decreased *as* the EROI declined steadily.

Although the EROI is an important metric, the truly critical element for the current civilization is the quality factor. A declining EROEI reveals that extraction of energy will be increasingly expensive and eventually, cost-prohibitive and the implications could be staggering [7]. Note that 'cost-prohibitive' energy extraction must be interpreted through the lens of actual geo-physical constraints; 'energy-prohibitive' energy extraction is more appropriate – after all, money (as a fungible proxy for cost) is a claim on energy. Ultimately the monetary costs of energy extraction will erode economic growth and eventually cap economic expansion. Strictly speaking, humanity could exploit all fossil fuels until the critical value of EROI=1 is reached (Figure 6). But modern industrial societies require an EROI of about 10 (this is an estimate); so one can expect 'economic convulsions' and various forms of social upheavals *well before* the hypothetical limit of EROI=1 is reached.

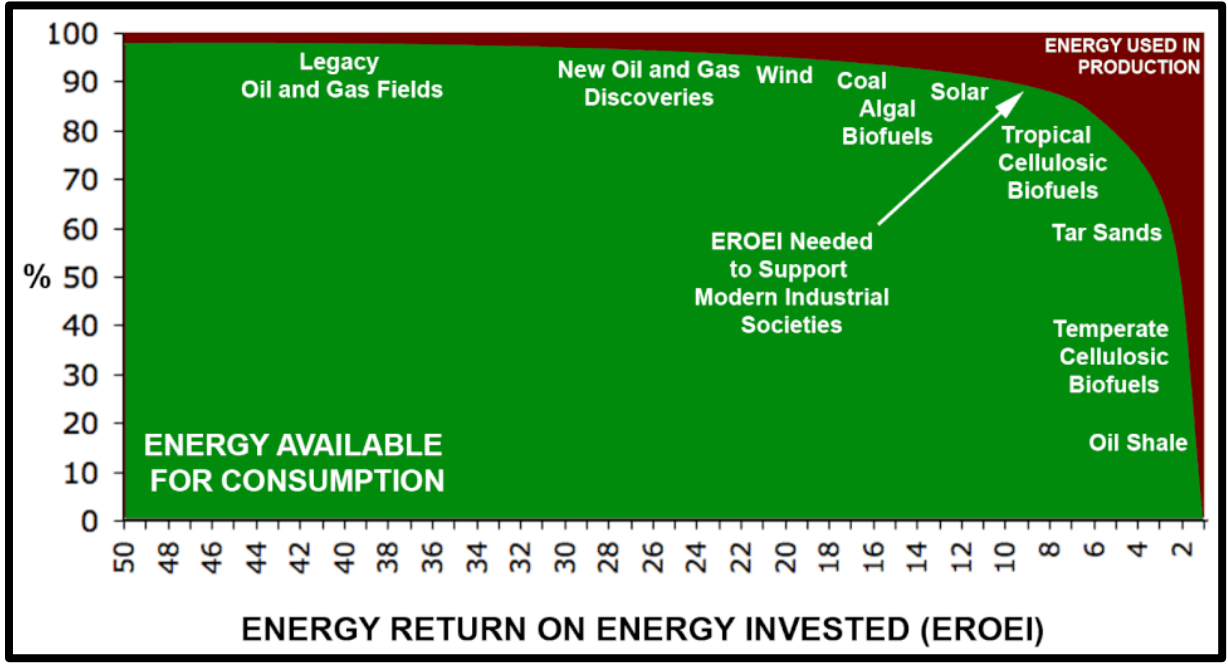

Figure 6. The energy cliff (adapted from Wikipedia)

The useful power for maintaining and growing the economic superorganism is the net power

$$P_{eff}(t) = \mathbb{Q}(\theta)\, P(t) \tag{10}$$

where the raw power is given by Equation (7). Equations (1),(5),(6),(10) do not form a closed set, as the dynamical evolution of the quality (as measured by the EROI) is yet undetermined. To do so, we need to quantify how fossil fuel resources are distributed in terms of quality. High-quality resources are rarer than resources of lower quality. We introduce the probability distribution of EROI, $f(\theta)$. The quantity of resources, $\Delta E$, with a quality between $\theta$ and $\theta + \Delta\theta$ is

$$\Delta E = E_T \int_{\theta}^{\theta+\Delta\theta} f(\theta)\, d\theta$$

Here $E_T$ is the total fossil fuel resources with an EROI greater than unity. Fossil fuel resources must provide more energy to the society that the energy cost to extract and process them (a point that we have already discussed above but worth re-iterating). The probability distribution must satisfy the normalization condition (constraint) of

$$\int_{1}^{\theta_{max}} f(\theta)d\theta = 1 \tag{11}$$

Here $\theta_{max}$ is the highest EROI achievable; since energy is needed to generate energy, this quantity cannot be infinite (in the early days of the fossil fuel exploitation values of the order of 100 were inferred by various authors). Figure 7 depicts our strategy to determine the dynamical evolution of the EROI.

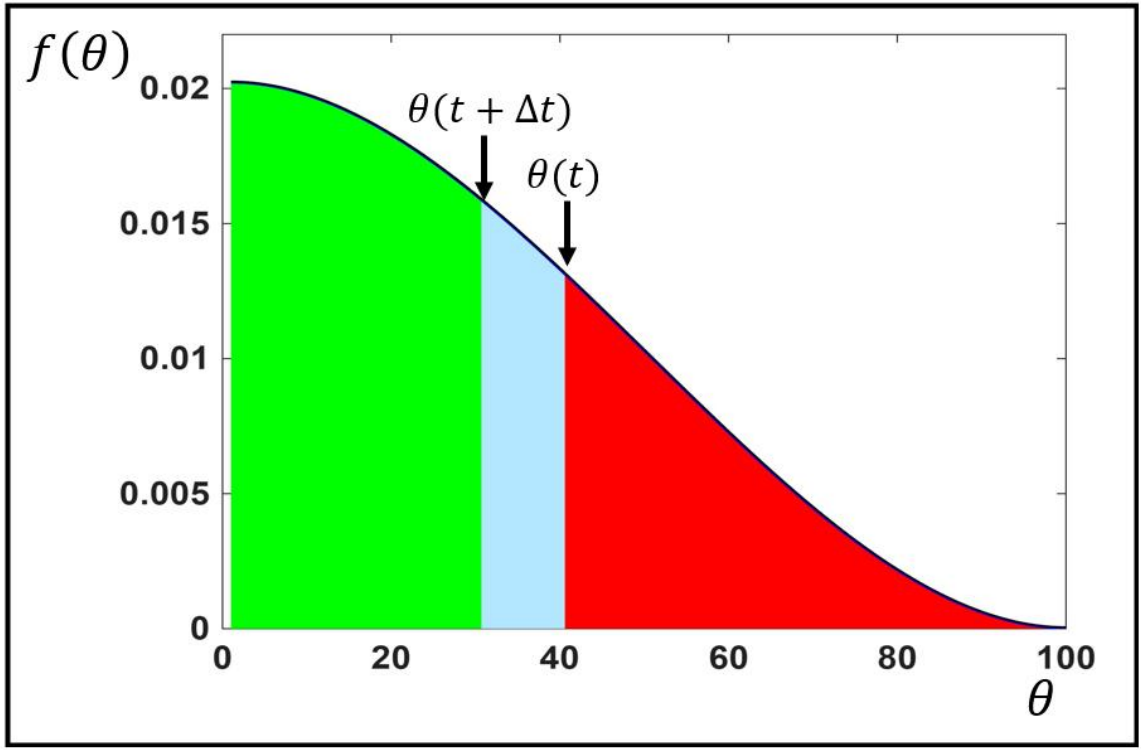

Figure 7: Strategy used to determine the dynamical equation for the EROI (denoted $\theta$). The low-hanging fruit paradigm applies: the easiest to extract resources are exploited first. As time proceeds, resources of lower quality, but more abundant, are tapped.

The cumulative energy extracted up to time $t$ is

$$E(t) = E_T \int_{\theta(t)}^{\theta_{max}} f(\theta)d\theta$$

whereas the cumulative energy extracted up to time $t + \Delta t$ is

$$E(t + \Delta t) = E_T \int_{\theta(t+\Delta t)}^{\theta_{max}} f(\theta)d\theta$$

Taking the difference between the above equations, and considering an arbitrarily small time increment $\Delta t$ we get

$$P(t)\Delta t \approx -f(\theta)E_T \Delta\theta$$

Dividing both sides of the above equation by the time increment and taking the limit $\Delta t \mapsto 0$, we obtain the equation for the dynamical evolution of the EROI:

$$\frac{d\theta}{dt} = -\frac{P(t)}{E_T f(\theta(t))} \qquad (12)$$

It is worth noting a couple of points. First, the greater the extraction rate of fossil fuels, the faster the drop in the quality of the extracted of resources: so a strategy of 'faster is better' is a sure race to the 'bottom of barrel'. Second, the actual shape of the distribution function $f(\theta)$ can, for a while, slow the ineluctable depletion of high-quality resources; the much applauded 'shale revolution' apparently provides 'new' resources to the civilization but shale oil's extraction is energy intensive (low EROI).

Equations (1),(5),(6),(10),(12) form a closed set of ordinary differential equations (ODEs) that must be solved numerically. The model is crude (pun not intended) but allows to capture the interplay of economic activity, power consumption, depletion and energy quality (Figure 8). We present numerical results in the next section.

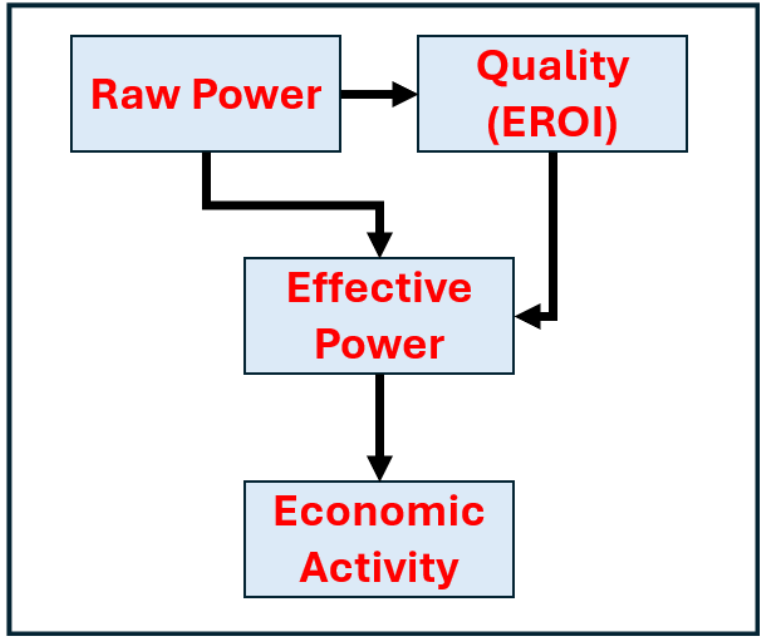

Figure 8: The driver for economic activity is effective power which is intertwined with raw power and its time-dependent quality (EROI).

Numerical Results

One critical element in the overall workflow of the power cliff model is the probability distribution function of resource quality. We suggest two different models for the distribution of EROI. The first model is called the polynomial model:

$$f(\theta) = A(1 - 3z^2 + 2z^3) \quad (13)$$

where $z = (\theta - 1)/(\theta_{max} - 1)$, $A = \dfrac{2}{\theta_{max}-1}$ and $\theta_{max}$ is the maximum EROI. The second model is called the Gaussian model:

$$f(\theta) = B(1 - z^2)\exp(-\nu z^2) \quad (14)$$

where $\nu$ is a free (positive) parameter; and $B = 1/[(\theta_{max} - 1)Q(\nu)]$ where

$$Q(\nu) = \frac{1}{2}\frac{\sqrt{\pi}}{\sqrt{\nu}}\mathrm{erf}(\sqrt{\nu})\left(1 - \frac{1}{2\nu}\right) + \frac{1}{\sqrt{\nu}}\exp(-\nu)$$

Both the above distributions satisfy the normalization condition of

$$\int_1^{\theta_{max}} f(\theta)d\theta = 1$$

The polynomial and Gaussian models are shown in Figure 9.

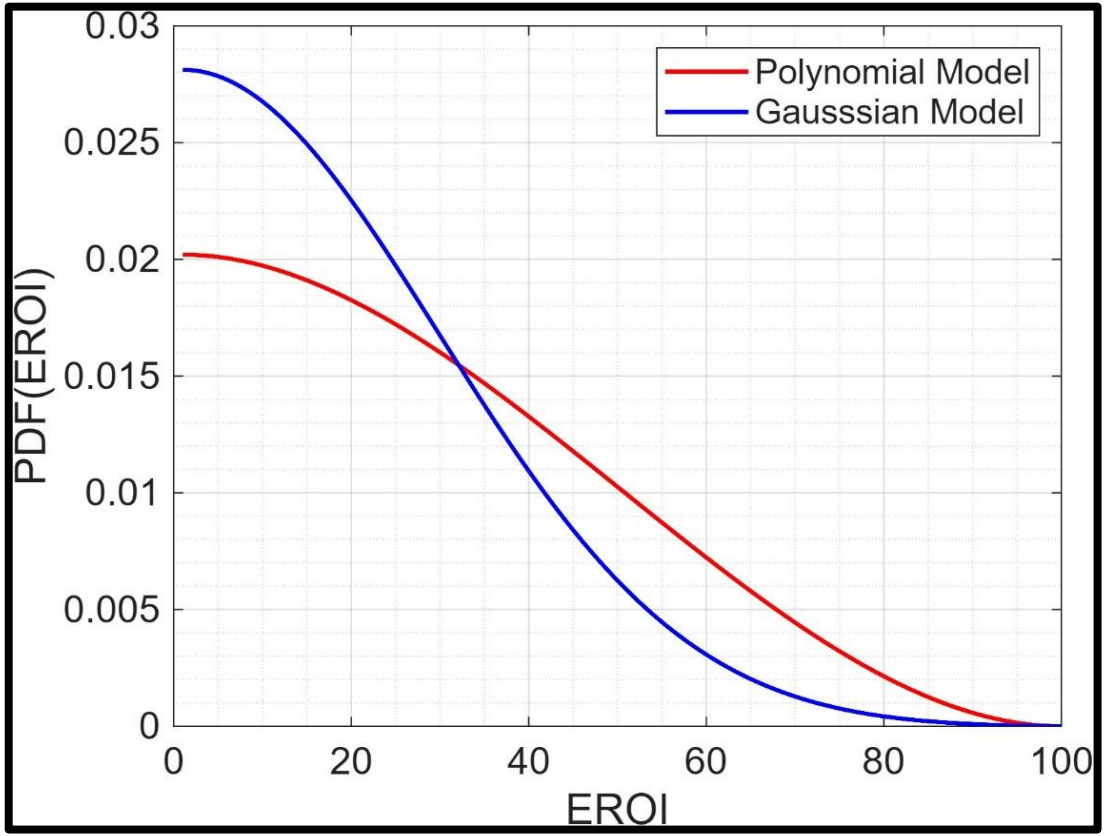

Figure 9: Comparison of the polynomial and Gaussian EROI distribution models for the case $\theta_{max} = 100$. The parameter for the Gaussian model is $\nu = 5$.

The set of ODEs (1),(5),(6),(10),(12) is solved using a fifth-order Runge-Kutta method [7]. A typical output is shown in Figure 10. The parameters are $\mu = 2$ and $x_c = 0.4$ for the economic growth rate [see Equation (5) and Figure 2]; and $\beta = 1.2$ ('high-technology' scenario; see Figure 4) and $\frac{\omega}{\gamma_M} = 0.5$ for the energy extraction rate [see Equation (6)].

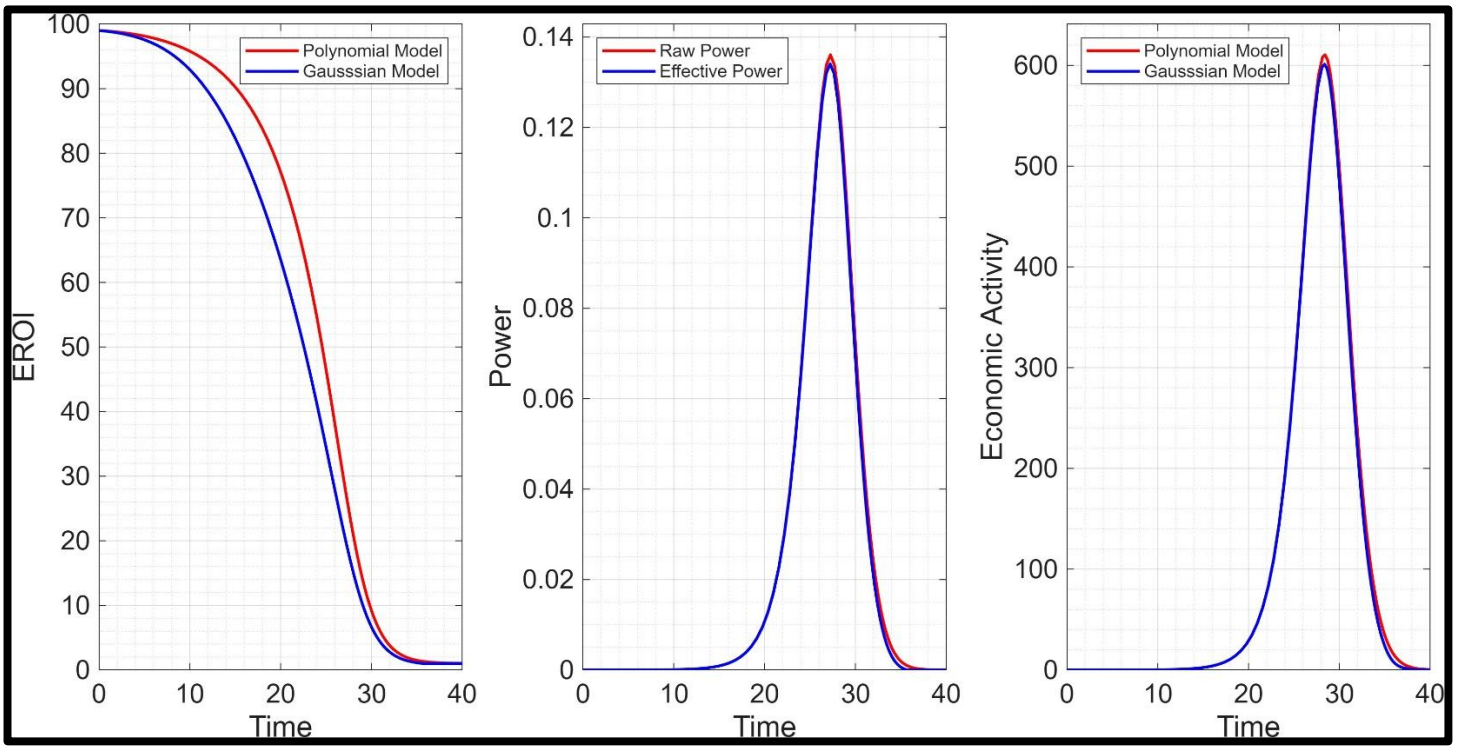

Figure 10. Typical dynamic evolution of the EROI (left panel), the (effective) power (middle panel) and the level of economic activity. Time is normalized to $1/\gamma_M$.

In a sense, this is an 'optimistic' scenario. First, we have assumed that the exponent $\beta$ [Equation (6)], which can be seen as proxy for technological prowess, is greater than one (M.K. Hubbert, in his landmark paper [5], suggested a value of $\beta = 1$ ). Second, the distributions of EROI [Equations (13),(14)] endow a significant amount of resources with high quality (Figure 9). Other parameters combinations have been tried but the overall dynamic is not significantly different from that shown in Figure 10. Of course, the power cliff model does not have predictive capabilities but it is a useful tool to better grasp the often-quoted remark due to E.F. Schumacher: 'Infinite growth of material consumption in a finite world is an impossibility' [9].